\documentclass[aps,prd,twocolumn,superscriptaddress,showpacs,nofootinbib]{revtex4-1}

\usepackage[utf8]{inputenc}

\usepackage{diagbox}

\usepackage{mathtools}
\usepackage{amsfonts}
\usepackage{mathrsfs}
\usepackage{bbm}
\usepackage{slashed}

\usepackage{graphicx}
\usepackage{color}
\usepackage{array}
\usepackage{esint}
\usepackage{placeins}
\usepackage{booktabs}
\usepackage{makecell}
\usepackage{epstopdf}
\usepackage[caption=false]{subfig}

\usepackage{xspace}
\usepackage{siunitx}
\usepackage{hyperref}
\usepackage[nameinlink]{cleveref}
\usepackage{appendix}

\usepackage{xifthen}
\usepackage{xcolor}
\hypersetup{
	colorlinks,
	linkcolor={red!75!black},
	citecolor={blue!75!black},
	urlcolor={blue!75!black}
}

\usepackage{comment}
\usepackage{dsfont}
\usepackage{tensor}
\usepackage{tabularx}
\usepackage{orcidlink}

\begin{document}


\title{Chiral Magnetic effect as the anomaly in the transverse axial vector Ward Identity}



\author{Fei Gao\,\orcidlink{0000-0001-5925-5110}}
\email{fei.gao@bit.edu.cn}
\affiliation{School of Physics, Beijing Institute of Technology, 100081 Beijing, China}

\author{Yi Lu}
\email[]{qwertylou@pku.edu.cn}
\affiliation{Department of Physics and State Key Laboratory of Nuclear Physics and Technology, Peking University, Beijing 100871, China}

\author{Minghui Ding}
\email{mhding@nju.edu.cn}
\affiliation{School of Physics, Nanjing University, Nanjing, Jiangsu 210093, China}
\affiliation{Institute for Nonperturbative Physics, Nanjing University, Nanjing, Jiangsu 210093, China}

\author{Xinyang Wang}
\email{wangxy@aust.edu.cn}
\affiliation{Center for Fundamental physics, School of Mechanics and Physics, Anhui University of Science and Technology, Huainan, Anhui 232001, China}

\author{Yuxin Liu}
\email{yxliu@pku.edu.cn}
\affiliation{Department of Physics and State Key Laboratory of
Nuclear Physics and Technology, Peking University, Beijing 100871,
China}
 \affiliation{Center for High Energy Physics, Peking
University, Beijing 100871, China}



\date{\today}

\begin{abstract}

Through analyzing the quark propagator under the magnetic field, we establish that the axial anomaly originates from an  additional Dirac structure  in quark propagator  induced by the magnetic field.  This Dirac structure also allows one to connect the axial anomaly with the topological properties of the system by checking the axial vector Ward identity.  For the tree level propagator,  we reproduce the  result of the anomalous axial current as in the Dirac Hamiltonian approach and kinetic theory. Particularly, we  confirm that the chiral magnetic  effect (CME) comes from the same term that is in charge of the  axial anomaly, specifically, as the anomaly of the transversal axial vector Ward Identity.  The identity guarantees that  the CME conductivity $C_{\rm CME}$ is a constant as $C_{\rm CME}=\frac{1}{2\pi^2}$, and is robust   against the temperature, chemical potential, magnetic field and also interaction. Finally, we verify this numerically by applying the full quark propagator under magnetic field calculated from the functional QCD methods.
 \end{abstract}


\maketitle



%

\section{Introduction}
The axial  anomaly  is an intriguing phenomenon in gauge field theory which has inspired extensive theoretical and experimental studies. Plenty of related anomalous effects have been discovered in different scenarios.  Under the magnetic field, there exist the chiral magnetic effect (CME) and chiral separation effect (CSE) which are evidently related to the axial anomaly~\cite{Son:2004tq,Metlitski:2005pr,Fukushima:2008xe,Kharzeev:2013ffa,Jiang:2016wve,Hou:2020zhb,Gao:2020vbh,Feng:2025yte}. The CME is particularly interesting, as the  conductivity of CSE is dependent on the temperature and quark mass~\cite{Brandt:2023wgf}, while that of CME is  widely believed to be robust against the temperature,  quark mass and  also magnetic field (up to a possible renormalization factor)~\cite{Yamamoto:2011gk,Kharzeev:2016sut, Brandt:2024wlw}.  This  distinctive  feature of CME   together with its anomalous nature has indicated its connection to the underlying topological structure of the system~\cite{Metlitski:2005pr,Fukushima:2008xe,Son:2012wh,Chen:2012ca,Chen:2014cla,Li:2014bha,Huang:2015oca, Kharzeev:2020jxw}.  

Generally speaking, the axial anomaly arises from the anomalous non conservation of the axial Ward identity.   The additional non conservation term in the Ward identity corresponds to the Adler-Bell-Jackiw triangle (ABJ) anomaly~\cite{Adler:1969gk,Bell:1969ts,Bardeen:1969md}, and can   equivalently be interpreted as the topological term~\cite{Fujikawa:1984pt,Son:2012wh}.  Notably,   the Ward identity can be largely extended through the local Lorentz transformation, which is the so called  transversal Ward identity~\cite{Kondo:1996xn,He:2000we,Qin:2013mta}. Parallel to the axial Ward identity,   the  extended transversal axial Ward identities  also  exhibit   the anomalous term.

 People have made a lot of efforts  on investigating the CME based on effective QCD models,  transport models, chiral kinetic theory and also hydrodynamics simulation~\cite{Fukushima:2008xe, Gynther:2010ed,Hou:2011ze,Gao:2012ix,Chen:2012ca,Chen:2014cla,Wu:2016dam,Hattori:2016emy,Deng:2016knn,Deng:2018dut,Zhao:2022grq,Jiang:2016wve,Huang:2018wdl,Shi:2019wzi,Kharzeev:2022hqz}.  In the work, we   convey a complete  investigation on    axial anomaly   based on  the exact form of  fermion propagator in the background of the  magnetic field from quantum field theory.  Particularly,  we classify the anomalous phenomenon into two different  categories, either  from the conventional Ward identity or the extended transversal Ward identity.  Crucially,  we elaborate that  the robustness of CME conductivity comes from the transversal axial Ward identity with anomaly, parallel to the topological term in the conventional axial Ward identity. 

\section{The quark propagator in  the magnetic field and the axial current}\label{sec:dispersion}
We start with quark propagator in the Ritus basis which is the complete basis under the magnetic field and   can be formulated as follows~\cite{Ferrer:2008dy,Watson:2013ghq,Ding:2025zqu}:
\begin{align}\label{eq:Ritus}
&\quad S^{-1}(p_\parallel,p_\perp)\notag\\
=&{}e^{-\frac{p_\perp^2}{|q_fB|}}
	\sum_{n=0}^\infty 2(-1)^{n}\biggl[2i \slashed{p}_\perp {Z_\parallel}(p_\parallel,n)  L_{n-1}^{1}\left(\alpha_\perp\right)\notag\\
	&-\Sigma^+(-i\slashed{p}_\parallel {Z_\parallel^+}(p_\parallel,n)+{M^+}(p_\parallel,n))L_{n-1}^{0}\left(\alpha_\perp\right)\notag\\
	&+\Sigma^-(-i\slashed{p}_\parallel {Z_\parallel^-}(p_\parallel,n)+{M^-}(p_\parallel,n))L_n^{0}\left(\alpha_\perp\right)\biggl]\,,
	\end{align}
	\begin{align}\label{eq:Ritus1}
	&\quad S^{}(p_\parallel,p_\perp ) \notag\\
	=&{}e^{-\frac{p_\perp^2}{|q_fB|}}
	\sum_{n=0}^\infty 2(-1)^{n} \biggl[2i \slashed{p}_\perp W_{Z_\parallel}(p_\parallel,n)  L_{n-1}^{1}\left(\alpha_\perp\right)\notag\\
	&-\Sigma^+(-i\slashed{p}_\parallel W_{Z_\parallel^+}(p_\parallel,n)+W_{M^+}(p_\parallel,n))L_{n-1}^{0}\left(\alpha_\perp\right)\notag\\
	&+\Sigma^-(-i\slashed{p}_\parallel W_{Z_\parallel^-}(p_\parallel,n)+W_{M^-}(p_\parallel,n))L_n^{0}\left(\alpha_\perp\right)\biggl]\,,
\end{align}
with $B$ the magnetic field strength and $q_f$ the electric charge of quark. $X=Z_\parallel^\pm,\,Z_\perp,\,M^\pm$  the dressing functions of the inverse quark propagator and $W_X$ the dressing functions of the quark propagator. Their relations are listed in the appendix~\ref{App:ritus}. The momentum   $p_\perp$  is the two dimensional transversal momentum $(0,\vec{p}_\perp,0) $ and  $p_\parallel=(\tilde p_0= \omega_m+i \mu_q, 0,0,p_3)$ with $\omega_m=(2m+1)\pi T$ the Matsubara frequency of quark and $\mu$ the quark chemical potential. $L_n^i(\alpha_\perp)$  is the generalized  Laguerre polynomials  with $\alpha_\perp=2 p_\perp^2/( q_fB)$. 
In the presence of magnetic field,   the spin-up and spin-down  components in the propagator are split, with the  spin projection matrices as:
\begin{align}
	\Sigma^\pm=\frac{1}{2}\left(\mathds{1}\pm \Sigma_3\right)\,,\quad \Sigma_3=i\gamma^1\gamma^2\,.
\end{align}
 
 If applying a tree level propagator which means that $Z_\parallel^\pm=Z_\perp=1$ and $M^\pm=m_f$, the inverse propagator in Eq.~\ref{eq:Ritus}  is the same as the free form as:
 \begin{align}
	S^{-1}= i\slashed{p}_\parallel+i\slashed{p}_\perp+ m_f.
	\end{align}
Therefore, there is no splitting between spin-up and down in the inverse quark propagator and only after considering the interaction that induces the splitting of $Z_\parallel^+,\, M^+$ and $Z_\parallel^-,\, M^-$ will lead to the splitting which is the anomalous Zeeman effect~\cite{Ferrer:2008dy,Ferrer:2009nq}. However, for the quark propagator,  the splitting exists even without interaction.
 The central element here that contributes to the splitting is the term related to the structure $ \Sigma_3 \slashed{p}_\parallel$ and $\Sigma_3 $ which is eventually related to the anomaly as we will elaborate below. For the tree level propagator, the respective coefficients of the propagator is written as:
\begin{align}\label{eq:tree1}
 &W_{Z_5}(p_\parallel,n)= W_{Z_\parallel^+}(p_\parallel,n+1)-W_{Z_\parallel^-}(p_\parallel,n) \notag\\
 =&-\frac{1}{p_\parallel^2+m_f^2+2n|q_fB|}+\frac{1}{p_\parallel^2+m_f^2+2(n+1)|q_fB|},\\
 &W_{M_5}(p_\parallel,n)= W_{M^+}(p_\parallel,n+1)-W_{M^-}(p_\parallel,n)\notag\\
 =&-\frac{m_f}{p_\parallel^2+m_f^2+2n|q_fB|}+\frac{m_f}{p_\parallel^2+m_f^2+2(n+1)|q_fB|}.\label{eq:tree2}
 \end{align}
 For massless particles,  the term $W_{M_5}$ vanishes and only  the term $W_{Z_5}$ contributes to the anomalous phenomena. 
 
 To see the relation between these terms and the anomaly,  one may   calculate the  axial current $J_5$  as:
\begin{align}
J_5=&<\bar{\psi}\gamma^3\gamma_5\psi>=-{\rm Tr}[\gamma^3\gamma_5 S]\notag\\
=& \frac{|q_f B| T}{4\pi} \sum_{m,n}\int \frac{dp_3}{2\pi} \left[- 4i \tilde p_0W_{Z_5}(p_\parallel,n)\right],
\end{align}
with  the momentum $p_\perp$ being integrated out and $m$ the index of Matsubara frequency,  $n$ the order of Laguerre polynomials.  

With the tree level propagator as in Eq.~\ref{eq:tree1}  and \ref{eq:tree2},  it is interesting to see that  the spin up and spin down components with $n>0$ cancel with each other in the current, and only  the number density at zeroth Landau level $n_0$ is left, which reads:

\begin{equation}
J_5= \frac{|q_f B| T}{4\pi}  n_0,\quad n_0=T\sum_m\int \frac{dp_3}{2\pi} \frac{-4 i\tilde p_0}{p^2_\parallel+m_f^2}. 
\end{equation}
It is not difficult to calculate the summation and the integration, which  gives:
\begin{eqnarray}\label{eq:J5}
J_5=&&\frac{|q_fB|  }{2\pi^2} \int_0^\infty dp_3 (\frac{1}{e^{\frac{\sqrt{p_3^2+m_f^2}-\mu_q}{T}}+1}-\frac{1}{e^{\frac{\sqrt{p_3^2+m_f^2}+\mu_q}{T}}+1})\notag\\
&&\xrightarrow{m_f\rightarrow 0}\frac{|q_fB|  }{2\pi^2}\mu_q.
\end{eqnarray}
This then  recovers the result from the Dirac Hamiltonian  approach and kinetic theory~\cite{Metlitski:2005pr, Fukushima:2008xe,Son:2012wh,Chen:2012ca,Chen:2014cla}.

Meanwhile, the situation for the asymmetry of number density  $n_5$ is subtle. One can still calculate $n_5$ directly from the free quark propagator as:
\begin{equation}\label{eq:n5}
n_5=<\bar{\psi}\gamma^0\gamma_5\psi>=\frac{|q_fB|  }{4\pi} T\sum_m \int \frac{dp_3}{2\pi}   \frac{ 4 ip_3}{p^2_\parallel+m_f^2}=0,
\end{equation}
The factor $p_3$ in the expression leads  the asymmetry of number density  to be vanishing if integrating out the momentum space.  However,  the momentum distribution of the asymmetry is actually not vanishing, and hence, the higher order fluctuations of $n_5$ are not vanishing.

Besides, one can consider an external source as $\mu_5 \bar{\psi}\gamma_0\gamma_5\psi$, it can be rewritten as $$\mu_5 \bar{\psi}\gamma_0\gamma_5\psi=\mu_5 \bar{\psi} (\Sigma^+-\Sigma^-)\gamma_3\psi.$$  Note that it induces the splitting of spin up and spin down quarks, which is equivalent to the splitting of $Z^\pm_\parallel$.  Consequently,  it is not   equivalent to  a simple inclusion of $\mu_5$  analog to  the chemical potential.  For the denominator of quark propagator, one has to replace $p^2_\parallel$ to $p^2_\parallel+\mu_5^2$.  For illustration, we again take the chiral limit for following derivations,  and  one has:
\begin{eqnarray}\label{eq:mu5}
n_5&&=\frac{|q_fB| }{2\pi^2} T\sum_{m,n} \int  dp_3   \bigg[ \frac{ i p_3+\mu_5}{p^2_\parallel +\mu_5^2+2n |q_fB|}\notag\\
&&- \frac{ ip_3-\mu_5}{p^2_\parallel +\mu_5^2+2(n+1) |q_fB|}\bigg]\notag\\
&&\xrightarrow{\rm{small} \, \mu_5}\frac{|q_fB|T \mu_5 }{\pi}\bigg[-\sum_{m=0}^\infty \frac{1}{(2m+1)\pi T} \notag\\
&&+ \sum_{n,m=0}^\infty \frac{2}{\sqrt{((2m+1)\pi T)^2+2n |q_fB|}}\bigg]. 
\end{eqnarray}
The summation in the last line of Eq.~\ref{eq:mu5} can be defined  by   analytic continuation of  the Hurwitz zeta function $\zeta(s,x)$ with $s=1/2$   from  its  integral representation,  which  for $\pi^2 T^2/(2|q_fB|)\gg 0$ can be expanded as:
\begin{eqnarray}\label{eq:n52}
 &&\sum_{n,m=0}^\infty \frac{2}{\sqrt{((2m+1)\pi T)^2+2n |q_fB|}}\\
 =&&\sum_{m=0}^\infty  (- \frac{2(2m+1)\pi T}{ |q_fB|}+\frac{1}{(2m+1)\pi T}+\rm{high\, order}).\notag
\end{eqnarray}
The second term in R.H.S. of Eq .~\ref{eq:n52} cancels the term in Eq.~\ref{eq:mu5} precisely, and only  the first term in Eq .~\ref{eq:n52}  is left in the number density asymmetry $n_5$, which is the Riemann zeta function and  can be calculated  through   analytic continuation   as:
\begin{eqnarray}\label{eq:mu52}
n_5= -4\mu_5 T^2 \left[\zeta(-1)+\frac{1}{2}\zeta(0)\right]=\frac{4}{3}\mu_5 T^2.
\end{eqnarray}

One can also apply the lowest Landau Level approximation for very large magnetic field. In this case only the term with $n=0$ is left, and $n_5$ becomes:
\begin{eqnarray}\label{eq:largeB}
n_5=\frac{ \mu_5 |q_fB| }{2\pi^2}\sum_m\frac{1}{m+\frac{1}{2}}.
\end{eqnarray}
The summation is the harmonic series and here is regularized by considering the replacement from $\zeta(1)$ to its Cauchy principle value $\frac{\zeta(1+\epsilon)+\zeta(1-\epsilon)}{2}=\gamma$ with $\gamma$ the Euler constant which   gives:
\begin{eqnarray}\label{eq:largeB2}
n_5=-\frac{ \mu_5 |q_fB| \psi^{(0)}{(\frac{1}{2}}) }{4\pi^2}\approx \frac{ \mu_5 |q_fB| }{2\pi^2},
\end{eqnarray}
where $\psi^{(0)}$ is the digamma function with $$-\psi^{(0)}(\frac{1}{2})=\gamma+\rm Log(4)\approx 2.$$
The inclusion of a uniform nonzero $\mu_5$ is subtle, and one may consider only the inclusion of a locally nonzero $\mu_5$.  Nevertheless, the above derivations show the connection between the Dirac structure $ \Sigma_3 \slashed{p}_\parallel$ and the axial current.  If there exists axial anomaly,  it  is then   generated  also   from this special Dirac structure induced by  the magnetic field background.

\section{The axial anomaly and the non conservation of axial current}\label{sec:WTI}
To see this more clearly,  one may check the current conservation   in chiral limit. First of all,  one defines the current in time and space as:
\begin{eqnarray}\label{eq:AVA}
&&J_5(x=(\tau,0,0,z))=\frac{|q_fB|  }{4\pi} T\sum_m \int  \frac{dp_3}{2\pi} \frac{-4i \tilde p_0 }{p_{||}^2}e^{i p\cdot  x}\notag\\
=&&\frac{|q_fB|}{2\pi^2}\int_0^\infty dp_3  \cos(p_3 z)\left[f(p_3)+f(-p_3)\right];\\
&&n_5(x=(\tau,0,0,z))=\frac{|q_fB|  }{4\pi}  T\sum_m \int  \frac{dp_3}{2\pi} \frac{ 4i p_3 }{p_{||}^2}e^{i p\cdot  x}\notag\\
=&&\frac{ |q_fB|}{2\pi^2}\int_0^\infty dp_3  \sin(p_3 z)\left[f(p_3)-f(-p_3)\right].
\end{eqnarray}
with $f(p_3)=\frac{e^{(p_3-\mu_q)\tau}}{e^{(p_3-\mu_q)\beta}+1}$. 
One can then check the axial anomaly with the above equations. In  imaginary time,  the current conservation can be expressed as:
\begin{eqnarray}\label{eq:anomalyR}
&&\int_{z>0} d^3 x \int^\beta_0 d\tau \partial_\mu J_5^\mu\\
=&&\frac{i|q_fB|  }{2\pi^2} \int dx dy  \int_0^{\infty}dz \int_0^{\infty} d p_3 \sin(p_3 z) \notag\\
&&\times(\frac{\mu_q}{p_3-\mu_q}\frac{e^{(p_3-\mu_q)\beta}-1}{e^{(p_3-\mu_q)\beta}+1}+\frac{\mu_q}{p_3+\mu_q}\frac{e^{(-p_3-\mu_q)\beta}-1}{e^{(-p_3-\mu_q)\beta}+1})\notag\\
=&&-\frac{ |q_fB| S}{ 2\pi^2}\sum_{\pm}[\psi^{(0)}(\frac{1}{2}\pm i\frac{\mu_q}{2\pi T})-\psi^{(0)}(\frac{1}{2})]\notag\\
 &&\xrightarrow{\rm small \,\mu}-\frac{ |q_fB| S}{ 2\pi^2}\frac{\mu_q^2 \psi^{(2)}(\frac{1}{2})}{4\pi^2T^2} \notag,
\end{eqnarray}
with $S=\int dx dy$, and the term is thus related to  the magnetic flux as counted in the topological term.   It is also possible to have the axial anomaly even at zero chemical potential if the Polyakov loop is not vanishing since the background field condensate  $\phi=\beta \bar{A}_0^a t^a$  is similar to the  imaginary chemical potential.  With similar procedure, one immediately gets:
\begin{eqnarray}\label{eq:anomalyI}
&&\int_{z>0} d^3 x \int^\beta_0 d\tau \partial_\mu J_5^\mu\notag\\
=&&-\frac{  |q_fB| S}{ 2\pi^2}\sum_{\pm}[\psi^{(0)}(\frac{1}{2}\pm\frac{\phi}{2\pi})-\psi^{(0)}(\frac{1}{2})].
\end{eqnarray}

In the current analysis,  only the positive $z$ direction is taken into account which corresponds  solely to the outgoing magnetic flux. If integrated out the whole space, the above integral vanishes  because the incoming and outgoing magnetic flux cancel   each other,  reflecting the interplay between the nontrivial topological structure and the  axial anomaly.    However, we emphasize here that  the anomaly itself is not intrinsically tied to the nontrivial topological structure. The nontrivial topology is the manifestation of the anomaly in the two point correlation function, while for three point correlation function, the anomaly can be simply manifested without considering the topology, which is the Adler-Bell-Jackiw anomaly. 

Note that a finite $\mu_5$ can also bring in the non conservation of the axial current. The finite $\mu_5$ leads to:
\begin{eqnarray}\label{eq:anomalyR2}
  \int^\beta_0 d\tau \partial_\mu J_5^\mu\sim  \mu_5  |q_fB|   e^{- \pi T |z|},\notag
\end{eqnarray}
which is non vanishing even with the whole space being integrated out.  This then implies that  the nontrivial topology can be represented by a non zero chiral chemical potential $\mu_5$. If a finite $\mu_5$ is generated spontaneously,  then there exists non trivial topology in the system.  However,   the non zero $\mu_5$ is  manifested  by the finite axial number density $n_5$ but does not necessarily link to the fluctuations which can still be nonzero at $\mu_5=0$, and hence,   the discussion on the topology requires a further distinction in the observables.

Besides,  we   emphasize that the anomalous term discussed above is the anomaly in the axial vector Ward identity,   whereas  in the  extended transverse Ward identity, one can also find the  anomalous term parallel to the anomaly in conventional Ward identity.  This anomaly appearing in the transverse Ward identity turns out to be the chiral magnetic effect as will be elaborated below.

\section{The chiral magnetic effect  as the anomaly in the transverse axial vector Ward identity}\label{sec:CME}
For further explanations, we first note that the axial  current conservation in  Eq.~\ref{eq:anomalyR}  can be  rewritten in a more general form as:
\begin{eqnarray}\label{eq:anomaly0}
  \nabla_\mu J_5^\mu=\tilde{F}_{\mu\nu}(\nabla_\mu\nabla_\nu-\nabla_\nu\nabla_\mu) \phi(x),
\end{eqnarray}
with $\tilde{F}_{\mu\nu}$ the field strength tensor coming from the magnetic field,  $\nabla_\nu$ the covariant derivative  which comes from the converting of  $p_\nu$ with the Fourier transformation basis $e^{i p x}$ and 
\begin{eqnarray}
\phi(x)= T\sum_m \int  \frac{dp_3}{2\pi^2} \frac{1 }{p_{||}^2+m_f^2}e^{i p\cdot  x}.
\end{eqnarray}
For instance, for $\mu_q$  and $m_f$ go to zero, one gets 
$$\phi(x)=\frac {1}{(4\pi^2)}\sum_{\pm}{\rm Arctanh}[e^{\pi T (\pm i\tau-|z|)}].$$
 Moreover,  if generalizing the covariant derivative to  the covariant derivative from the gauge field as $\nabla_\mu\rightarrow D_\mu$, one  has:
\begin{eqnarray}\label{eq:anomaly1}
  \nabla_\mu J_5^\mu=\tilde{F}_{\mu\nu}F_{\mu\nu} \phi(x),
\end{eqnarray}
with $[D_\mu,D_\nu]=F_{\mu\nu}$ coming from the interaction of high order. This is nothing but the on-shell axial Ward identity with the anomalous term.  As discussed above, if considering only the positive $z$ direction, the anomaly is non vanishing, which offers a simple case for that  the nontrivial topology manifests the anomaly. Another case is to regard  the function $\phi(x)$  as the solitonic solutions, where the anomaly gives the topological charge of the soliton field as have been widely studied~\cite{Goldstone:1981kk,Son:2007ny}.

 The CME cannot be directly related to the Ward identity, instead, it is related to the anomaly in the transverse Ward identity of the axial vector, which is a parallel relation to the axial  vector Ward identity and can be derived after applying the Lorentz transformation to the Ward identity. To see its connection with the transverse Ward identity, one may directly consider the chiral magnetic conductivity defined as:
\begin{eqnarray}\label{eq:CME}
&&|q_f B| C^{}_{\rm CME}=\frac{\partial J_3}{\partial \mu_5}{\bigg |}_{\mu_5=0}\\
=&&\int d^4 y < \bar{\psi}_f(x)\gamma_3\psi_f(x)  \bar{\psi}_f(y)\gamma_0\gamma_5\psi_f(y) >,\notag
\end{eqnarray}
if one puts a tree level propagator in the definition, the conductivity  can be directly computed as:
\begin{eqnarray}\label{eq:CME1}
C^{0}_{\rm CME}=&&\frac{T}{\pi} \int \frac{d  p_3}{ 2\pi} \sum_{m,n}     [\frac{ \tilde{p}_0^2-p_3^2+m_f^2}{(p_\parallel^2+m_f^2+2n|q_f B|)^2}\notag\\
&&- \frac{ \tilde{p}_0^2-p_3^2+m_f^2}{(p_\parallel^2+m_f^2+2(n+1)|q_f B|)^2}]\notag\\
=&& \frac{1}{2\pi^2}  \frac{\sqrt{1 - (m_f/\epsilon)^2} \sinh (\epsilon/T)}{\cosh (\epsilon/T)+\cosh(\mu/T)}\biggr|_{\epsilon=m_f}^{\epsilon=+\infty} \notag \\
=&& \frac{1}{2\pi^2},
\end{eqnarray}
where again $\tilde{p}_0 = (2m+1)\pi T + i \mu_q$ is the quark momentum at finite temperature and chemical potential.
One gets a striking result that the above integral  is precisely equal to $\frac{1}{2\pi^2}$ for any temperature, chemical potential and magnetic field. Note that  the integral is  well defined if one applies the real time formula and computes  it in the time-like region\footnote{In the real time formula, one needs to carefully deal with the ordering of  the integral $\int dp_0 dp_3$ since the pole position is at $p_0^2-p_3^2-m_f^2=0$. If one integrates $p_3$ direction firstly,  the integral of $p_0$ should be separated into two parts with the regime $p^2_0< m_f^2$ and  $p^2_0\geq m_f^2$ together with different parametrization of $p_3$ since in one regime, the momentum $p_3$ is real while in the other,  is imaginary. }.    Such a result indicates that the CME should be able to  related to some symmetry relations such as  Ward identities. Indeed, Eq.~\ref{eq:CME}  can be generalized in a covariant form as:
\begin{align}\label{eq:AvWTI}
&\int d^4 x \epsilon^{\mu\nu\rho\lambda}\bigg( \bar{\psi}(x)\sigma_{\rho\lambda^\prime}\gamma_5\nabla_{\lambda^\prime}\psi(x)\bar J_\lambda(y)+ m_f{J}^5_{\rho}(x) \bar J_\lambda(y)\bigg) \notag\\
&=\frac{m_f}{2\pi^2}F^{\mu\nu},
\end{align}
where  $ {J}^5_\lambda(x)=\bar{\psi}(x){\gamma}_\lambda\gamma_5\psi(x)$ and $\bar{J}_\rho(x)=\bar{\psi}(x)\bar{\gamma}_\rho\psi(x)$ with $\bar{\gamma}_\lambda=(0,\gamma_3)$. Such a choice of Eq.~\ref{eq:AvWTI}    gives   a special case of  Ref.~\cite{Kondo:1996xn}  as the    transverse Ward identity of axial vector current with anomalous term.  The R.H.S. of Eq.~\ref{eq:AvWTI} is the anomalous term which stands without the nontrivial topological structure.  The CME identity  in Eq~\ref{eq:CME1}  is obtained with the assist of another transverse Ward identity as:
\begin{align}\label{eq:AvWTI1}
&\int_x\epsilon^{\mu\nu\rho\lambda}  \bar{\psi}(x) [\gamma_\rho\gamma_5,\!\slashed{\nabla}]\psi(x)\tilde J_\lambda(y)\!- \!\bar{\psi}(x) [\sigma_{\mu\nu},\!\slashed\nabla]\psi(x) J(y)\notag\\
&\equiv\int d^4 x\epsilon^{\mu\nu\rho\lambda}  \bar{\psi}(x)\sigma_{\rho\lambda^\prime}\gamma_5\nabla_{\lambda^\prime}\psi(x)\bar J_\lambda(y)=0,
\end{align}
with $ {J}(x)=\bar{\psi}(x)\psi(x)$ and $\tilde{J}_\rho(x)=\bar{\psi}(x)\tilde{\gamma}_\rho\psi(x)$ with $\tilde{\gamma}_\lambda=(\gamma_0,0)$. The combination of Eq.~\ref{eq:AvWTI} and \ref{eq:AvWTI1} gives the identity for CME conductivity which then
  assures  that once the axial symmetry is preserved up to the anomaly, the conductivity of CME is unchanged as $C^{}_{\rm CME}=\frac{1}{2\pi^2}$, against the temperature, chemical potential, magnetic field and also the interaction. The CME current can be then written as:
\begin{equation}
J_3=\mu_5 |q_fB| \frac{1}{2\pi^2}= 
\left\{
\begin{aligned}\frac{3 |q_fB| }{8\pi^2T^2} n_5, \quad&  |q_f B|\ll T^2\\
 n_5, \quad &  |q_f B|\gg  T^2
\end{aligned}
\right.
\end{equation}
with Eq.~\ref{eq:mu52} being applied to get the relation between $J_3$ and $n_5$.   In the following, we will apply the full quark propagator calculated from Dyson Schwinger equations  to show  the validity  of the constant CME conductivity numerically.

\section{Numerical results for CME conductivity}\label{sec:results}
The conductivity of CME with the full propagator can be formulated as:
\begin{eqnarray}
&&   C^{}_{\rm CME}
=T \sum_{m,n} \int \frac{d  p_3}{ 2\pi^2}   [W^2_{M^-}(p_\parallel,n)-W^2_{M^+}(p_\parallel,n+1)\notag\\
&&+(\tilde p_0^2-p_3^2)(W^2_{Z_\parallel^-}(p_\parallel,n)-W^2_{Z_\parallel^+}(p_\parallel,n+1))]. 
\end{eqnarray}
The full quark propagator can be obtained after solving the Dyson-Schwinger equation at finite magnetic field. The detailed formula of the gap equation is listed in the appendix~\ref{App:ritus}, and here we simply give the results of the CME conductivity after incorporating the full propagator at different temperature and chemical potential. 

\begin{figure}[t]
  \centering
  \includegraphics[width=1.\columnwidth]{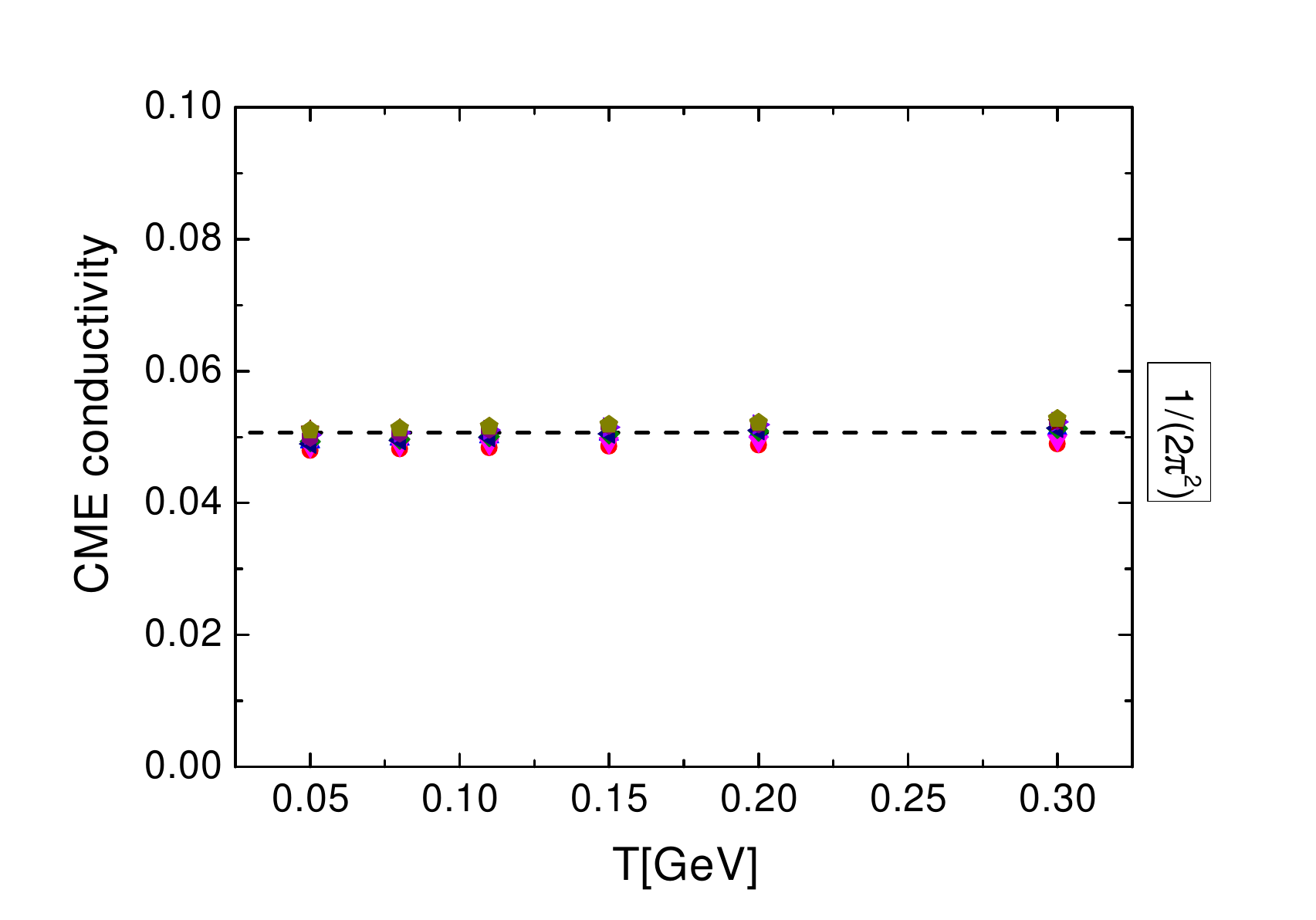}
  \caption{The conductivity of CME as a function of temperature, for different chemical potential in the range of $\mu_q\in[0,150]$~MeV and magnetic field of $q_f B\in[0, 2]~$GeV$^2$ represented by the points. Within the error bars, the CME conductivity remains the same as $C_{\rm CME}=\frac{1}{2\pi^2}$ for all  $T$, $\mu_B$ and $q_f B$. }\label{fig:CME}
\end{figure}

There contains a logarithm divergence in  the conductivity,  which can be  isolated  as  a term similar to the  tree level  propagator contribution.   This tree level term denoting as $ C^{\rm f}_{\rm CME}$ is written as:
\begin{eqnarray}
&& C^{\rm f}_{\rm CME}\\
=&&T \sum_{m,n} \int \frac{d  p_3}{ 2\pi^2}     [\frac{ (Z_\parallel^+Z_\parallel^-)(\tilde p_0^2-p_3^2)+M^+M^-}{(Z_\parallel^+Z_\parallel^-p_\parallel^2+M^+M^-+2n|q_f B| Z^2_\perp)^2}\notag\\
&&- \frac{ (Z_\parallel^+Z_\parallel^-)(\tilde p_0^2-p_3^2)+M^+M^-}{(Z_\parallel^+Z_\parallel^-p_\parallel^2+M^+M^-+2(n+1)|q_f B|Z^2_\perp)^2}]\notag\\
=&&C^{\rm  0}_{\rm CME}+\delta C^{}_{\rm CME}=\frac{1}{2\pi^2}+\delta C^{}_{\rm CME}.\notag
\end{eqnarray}
The divergence  can be avoided by numerically subtracting $C^{\rm  0}_{\rm CME}$ and calculating only the difference $\delta C^{}_{\rm CME}$.  The other term in the conductivity is induced purely by the self energy,  in specific,  by the splitting  of the coefficients in the inverse quark propagator $Z_\parallel^\pm$ and $M^\pm$. Such an interaction term does not contain any divergence and turns out to be $\mathcal{O}(10^{-2})$ smaller than the tree level term. 

The above separation has successfully handled the possible convergence in  the CME conductivity  and after then, the CME conductivity can be numerically calculated with a full quark propagator, which is   depicted in Fig.~\ref{fig:CME}. 
The conductivity is consistent with $\frac{1}{2\pi^2}$  for any temperature and chemical potential, as well as the magnetic field with the deviation  less than $5\%$.

\section{Summary}\label{sec:summary}

We systematically investigate the anomaly phenomena in the presence of  magnetic field starting from  the two point correlation function, i.e., the quark propagator   in the quantum field theory.  We demonstrate  that the magnetic field induces the  Dirac structure in quark propagator $\Sigma_3 \slashed{p}_\perp$, which explicitly  splits the spin up and down components of quark and  gives rise to the anomaly effects.   Such  an  Ab-initio method gives a unifying description for the anomalous properties under magnetic field, which enables one to classify the   anomaly phenomena. We discover that the CME originates from the anomaly of the transversal Ward identity, which  parallels  to the anomaly of the conventional Ward identity including the triangle anomaly and also the topological properties.  The transversal Ward identity yields the CME conductivity to be $\frac{1}{2\pi^2}$,  independent of  temperature, chemical potential and also interaction. This work bridges microscopic Dirac structure with macroscopic anomalous transport, resolving longstanding ambiguities in the interplay of anomaly, Ward identities, and observable effects.

Our results also indicate that  the relation between the anomaly and the topology is subtle. The non trivial topological structure is not directly related to all the anomalous properties, and rather, it is  one of the manifesting ways for the anomaly.  For instance,  the fluctuation of $n_5$ and consequently of the CME current can be  non vanishing, as the result of anomaly, but with no requirements on the non trivial topology.  The non trivial topology can affect the observables as it is equivalent to the inclusion of  the effective chiral chemical potential.  The non vanishing chiral chemical potential can lead to non zero $n_5$ and also some additional corrections on the CME current. Nevertheless,   the non conservation of  the   number density asymmetry as a function of time is the definite  observable  to connect with the non trivial topology.   This  distinction imposes  a strong constraint on  the possible measurements to   detect the topological structure. 
\begin{acknowledgments}
FG   thanks Haojie Xu, Yicheng Feng, Shuzhe Shi and Shi Pu for very helpful discussions.
The work is  supported by the National  Science Foundation of China under Grants  No. 12305134, 12247107, 12175007 and 12505101.   The work of X.W. is supported by YJ20240001 from Anhui University of Science and Technology.
\end{acknowledgments}

\bibliography{bibreferences}

\begin{thebibliography}{42}%
\makeatletter
\providecommand \@ifxundefined [1]{%
 \@ifx{#1\undefined}
}%
\providecommand \@ifnum [1]{%
 \ifnum #1\expandafter \@firstoftwo
 \else \expandafter \@secondoftwo
 \fi
}%
\providecommand \@ifx [1]{%
 \ifx #1\expandafter \@firstoftwo
 \else \expandafter \@secondoftwo
 \fi
}%
\providecommand \natexlab [1]{#1}%
\providecommand \enquote  [1]{``#1''}%
\providecommand \bibnamefont  [1]{#1}%
\providecommand \bibfnamefont [1]{#1}%
\providecommand \citenamefont [1]{#1}%
\providecommand \href@noop [0]{\@secondoftwo}%
\providecommand \href [0]{\begingroup \@sanitize@url \@href}%
\providecommand \@href[1]{\@@startlink{#1}\@@href}%
\providecommand \@@href[1]{\endgroup#1\@@endlink}%
\providecommand \@sanitize@url [0]{\catcode `\\12\catcode `\$12\catcode
  `\&12\catcode `\#12\catcode `\^12\catcode `\_12\catcode `\%12\relax}%
\providecommand \@@startlink[1]{}%
\providecommand \@@endlink[0]{}%
\providecommand \url  [0]{\begingroup\@sanitize@url \@url }%
\providecommand \@url [1]{\endgroup\@href {#1}{\urlprefix }}%
\providecommand \urlprefix  [0]{URL }%
\providecommand \Eprint [0]{\href }%
\providecommand \doibase [0]{http://dx.doi.org/}%
\providecommand \selectlanguage [0]{\@gobble}%
\providecommand \bibinfo  [0]{\@secondoftwo}%
\providecommand \bibfield  [0]{\@secondoftwo}%
\providecommand \translation [1]{[#1]}%
\providecommand \BibitemOpen [0]{}%
\providecommand \bibitemStop [0]{}%
\providecommand \bibitemNoStop [0]{.\EOS\space}%
\providecommand \EOS [0]{\spacefactor3000\relax}%
\providecommand \BibitemShut  [1]{\csname bibitem#1\endcsname}%
\let\auto@bib@innerbib\@empty
\bibitem [{\citenamefont {Son}\ and\ \citenamefont
  {Zhitnitsky}(2004)}]{Son:2004tq}%
  \BibitemOpen
  \bibfield  {author} {\bibinfo {author} {\bibfnamefont {D.~T.}\ \bibnamefont
  {Son}}\ and\ \bibinfo {author} {\bibfnamefont {A.~R.}\ \bibnamefont
  {Zhitnitsky}},\ }\href {\doibase 10.1103/PhysRevD.70.074018} {\bibfield
  {journal} {\bibinfo  {journal} {Phys. Rev. D}\ }\textbf {\bibinfo {volume}
  {70}},\ \bibinfo {pages} {074018} (\bibinfo {year} {2004})},\ \Eprint
  {http://arxiv.org/abs/hep-ph/0405216} {arXiv:hep-ph/0405216} \BibitemShut
  {NoStop}%
\bibitem [{\citenamefont {Metlitski}\ and\ \citenamefont
  {Zhitnitsky}(2005)}]{Metlitski:2005pr}%
  \BibitemOpen
  \bibfield  {author} {\bibinfo {author} {\bibfnamefont {M.~A.}\ \bibnamefont
  {Metlitski}}\ and\ \bibinfo {author} {\bibfnamefont {A.~R.}\ \bibnamefont
  {Zhitnitsky}},\ }\href {\doibase 10.1103/PhysRevD.72.045011} {\bibfield
  {journal} {\bibinfo  {journal} {Phys. Rev. D}\ }\textbf {\bibinfo {volume}
  {72}},\ \bibinfo {pages} {045011} (\bibinfo {year} {2005})},\ \Eprint
  {http://arxiv.org/abs/hep-ph/0505072} {arXiv:hep-ph/0505072} \BibitemShut
  {NoStop}%
\bibitem [{\citenamefont {Fukushima}\ \emph {et~al.}(2008)\citenamefont
  {Fukushima}, \citenamefont {Kharzeev},\ and\ \citenamefont
  {Warringa}}]{Fukushima:2008xe}%
  \BibitemOpen
  \bibfield  {author} {\bibinfo {author} {\bibfnamefont {K.}~\bibnamefont
  {Fukushima}}, \bibinfo {author} {\bibfnamefont {D.~E.}\ \bibnamefont
  {Kharzeev}}, \ and\ \bibinfo {author} {\bibfnamefont {H.~J.}\ \bibnamefont
  {Warringa}},\ }\href {\doibase 10.1103/PhysRevD.78.074033} {\bibfield
  {journal} {\bibinfo  {journal} {Phys. Rev. D}\ }\textbf {\bibinfo {volume}
  {78}},\ \bibinfo {pages} {074033} (\bibinfo {year} {2008})},\ \Eprint
  {http://arxiv.org/abs/0808.3382} {arXiv:0808.3382 [hep-ph]} \BibitemShut
  {NoStop}%
\bibitem [{\citenamefont {Kharzeev}(2014)}]{Kharzeev:2013ffa}%
  \BibitemOpen
  \bibfield  {author} {\bibinfo {author} {\bibfnamefont {D.~E.}\ \bibnamefont
  {Kharzeev}},\ }\href {\doibase 10.1016/j.ppnp.2014.01.002} {\bibfield
  {journal} {\bibinfo  {journal} {Prog. Part. Nucl. Phys.}\ }\textbf {\bibinfo
  {volume} {75}},\ \bibinfo {pages} {133} (\bibinfo {year} {2014})},\ \Eprint
  {http://arxiv.org/abs/1312.3348} {arXiv:1312.3348 [hep-ph]} \BibitemShut
  {NoStop}%
\bibitem [{\citenamefont {Jiang}\ \emph {et~al.}(2018)\citenamefont {Jiang},
  \citenamefont {Shi}, \citenamefont {Yin},\ and\ \citenamefont
  {Liao}}]{Jiang:2016wve}%
  \BibitemOpen
  \bibfield  {author} {\bibinfo {author} {\bibfnamefont {Y.}~\bibnamefont
  {Jiang}}, \bibinfo {author} {\bibfnamefont {S.}~\bibnamefont {Shi}}, \bibinfo
  {author} {\bibfnamefont {Y.}~\bibnamefont {Yin}}, \ and\ \bibinfo {author}
  {\bibfnamefont {J.}~\bibnamefont {Liao}},\ }\href {\doibase
  10.1088/1674-1137/42/1/011001} {\bibfield  {journal} {\bibinfo  {journal}
  {Chin. Phys. C}\ }\textbf {\bibinfo {volume} {42}},\ \bibinfo {pages}
  {011001} (\bibinfo {year} {2018})},\ \Eprint
  {http://arxiv.org/abs/1611.04586} {arXiv:1611.04586 [nucl-th]} \BibitemShut
  {NoStop}%
\bibitem [{\citenamefont {Hou}\ \emph {et~al.}(2021)\citenamefont {Hou},
  \citenamefont {Huang}, \citenamefont {Liao}, \citenamefont {Shi},\ and\
  \citenamefont {Zhang}}]{Hou:2020zhb}%
  \BibitemOpen
  \bibfield  {author} {\bibinfo {author} {\bibfnamefont {D.}~\bibnamefont
  {Hou}}, \bibinfo {author} {\bibfnamefont {A.}~\bibnamefont {Huang}}, \bibinfo
  {author} {\bibfnamefont {J.}~\bibnamefont {Liao}}, \bibinfo {author}
  {\bibfnamefont {S.}~\bibnamefont {Shi}}, \ and\ \bibinfo {author}
  {\bibfnamefont {H.}~\bibnamefont {Zhang}},\ }\href {\doibase
  10.1016/j.nuclphysa.2020.121971} {\bibfield  {journal} {\bibinfo  {journal}
  {Nucl. Phys. A}\ }\textbf {\bibinfo {volume} {1005}},\ \bibinfo {pages}
  {121971} (\bibinfo {year} {2021})},\ \Eprint
  {http://arxiv.org/abs/2004.00569} {arXiv:2004.00569 [nucl-th]} \BibitemShut
  {NoStop}%
\bibitem [{\citenamefont {Gao}\ \emph {et~al.}(2020)\citenamefont {Gao},
  \citenamefont {Ma}, \citenamefont {Pu},\ and\ \citenamefont
  {Wang}}]{Gao:2020vbh}%
  \BibitemOpen
  \bibfield  {author} {\bibinfo {author} {\bibfnamefont {J.-H.}\ \bibnamefont
  {Gao}}, \bibinfo {author} {\bibfnamefont {G.-L.}\ \bibnamefont {Ma}},
  \bibinfo {author} {\bibfnamefont {S.}~\bibnamefont {Pu}}, \ and\ \bibinfo
  {author} {\bibfnamefont {Q.}~\bibnamefont {Wang}},\ }\href {\doibase
  10.1007/s41365-020-00801-x} {\bibfield  {journal} {\bibinfo  {journal} {Nucl.
  Sci. Tech.}\ }\textbf {\bibinfo {volume} {31}},\ \bibinfo {pages} {90}
  (\bibinfo {year} {2020})},\ \Eprint {http://arxiv.org/abs/2005.10432}
  {arXiv:2005.10432 [hep-ph]} \BibitemShut {NoStop}%
\bibitem [{\citenamefont {Feng}\ \emph {et~al.}(2025)\citenamefont {Feng},
  \citenamefont {Voloshin},\ and\ \citenamefont {Wang}}]{Feng:2025yte}%
  \BibitemOpen
  \bibfield  {author} {\bibinfo {author} {\bibfnamefont {Y.}~\bibnamefont
  {Feng}}, \bibinfo {author} {\bibfnamefont {S.~A.}\ \bibnamefont {Voloshin}},
  \ and\ \bibinfo {author} {\bibfnamefont {F.}~\bibnamefont {Wang}},\ }\href
  {\doibase 10.1103/xk4f-wm9s} {\bibfield  {journal} {\bibinfo  {journal}
  {Phys. Rev. Res.}\ }\textbf {\bibinfo {volume} {7}},\ \bibinfo {pages}
  {031001} (\bibinfo {year} {2025})},\ \Eprint
  {http://arxiv.org/abs/2502.09742} {arXiv:2502.09742 [nucl-ex]} \BibitemShut
  {NoStop}%
\bibitem [{\citenamefont {Brandt}\ \emph
  {et~al.}(2024{\natexlab{a}})\citenamefont {Brandt}, \citenamefont
  {Endr{\H{o}}di}, \citenamefont {Garnacho-Velasco},\ and\ \citenamefont
  {Mark{\'o}}}]{Brandt:2023wgf}%
  \BibitemOpen
  \bibfield  {author} {\bibinfo {author} {\bibfnamefont {B.~B.}\ \bibnamefont
  {Brandt}}, \bibinfo {author} {\bibfnamefont {G.}~\bibnamefont
  {Endr{\H{o}}di}}, \bibinfo {author} {\bibfnamefont {E.}~\bibnamefont
  {Garnacho-Velasco}}, \ and\ \bibinfo {author} {\bibfnamefont
  {G.}~\bibnamefont {Mark{\'o}}},\ }\href {\doibase 10.1007/JHEP02(2024)142}
  {\bibfield  {journal} {\bibinfo  {journal} {JHEP}\ }\textbf {\bibinfo
  {volume} {02}},\ \bibinfo {pages} {142} (\bibinfo {year}
  {2024}{\natexlab{a}})},\ \Eprint {http://arxiv.org/abs/2312.02945}
  {arXiv:2312.02945 [hep-lat]} \BibitemShut {NoStop}%
\bibitem [{\citenamefont {Yamamoto}(2011)}]{Yamamoto:2011gk}%
  \BibitemOpen
  \bibfield  {author} {\bibinfo {author} {\bibfnamefont {A.}~\bibnamefont
  {Yamamoto}},\ }\href {\doibase 10.1103/PhysRevLett.107.031601} {\bibfield
  {journal} {\bibinfo  {journal} {Phys. Rev. Lett.}\ }\textbf {\bibinfo
  {volume} {107}},\ \bibinfo {pages} {031601} (\bibinfo {year} {2011})},\
  \Eprint {http://arxiv.org/abs/1105.0385} {arXiv:1105.0385 [hep-lat]}
  \BibitemShut {NoStop}%
\bibitem [{\citenamefont {Kharzeev}\ \emph {et~al.}(2017)\citenamefont
  {Kharzeev}, \citenamefont {Stephanov},\ and\ \citenamefont
  {Yee}}]{Kharzeev:2016sut}%
  \BibitemOpen
  \bibfield  {author} {\bibinfo {author} {\bibfnamefont {D.~E.}\ \bibnamefont
  {Kharzeev}}, \bibinfo {author} {\bibfnamefont {M.~A.}\ \bibnamefont
  {Stephanov}}, \ and\ \bibinfo {author} {\bibfnamefont {H.-U.}\ \bibnamefont
  {Yee}},\ }\href {\doibase 10.1103/PhysRevD.95.051901} {\bibfield  {journal}
  {\bibinfo  {journal} {Phys. Rev. D}\ }\textbf {\bibinfo {volume} {95}},\
  \bibinfo {pages} {051901} (\bibinfo {year} {2017})},\ \Eprint
  {http://arxiv.org/abs/1612.01674} {arXiv:1612.01674 [hep-ph]} \BibitemShut
  {NoStop}%
\bibitem [{\citenamefont {Brandt}\ \emph
  {et~al.}(2024{\natexlab{b}})\citenamefont {Brandt}, \citenamefont
  {Endr{\H{o}}di}, \citenamefont {Garnacho-Velasco},\ and\ \citenamefont
  {Mark{\'o}}}]{Brandt:2024wlw}%
  \BibitemOpen
  \bibfield  {author} {\bibinfo {author} {\bibfnamefont {B.~B.}\ \bibnamefont
  {Brandt}}, \bibinfo {author} {\bibfnamefont {G.}~\bibnamefont
  {Endr{\H{o}}di}}, \bibinfo {author} {\bibfnamefont {E.}~\bibnamefont
  {Garnacho-Velasco}}, \ and\ \bibinfo {author} {\bibfnamefont
  {G.}~\bibnamefont {Mark{\'o}}},\ }\href {\doibase 10.1007/JHEP09(2024)092}
  {\bibfield  {journal} {\bibinfo  {journal} {JHEP}\ }\textbf {\bibinfo
  {volume} {09}},\ \bibinfo {pages} {092} (\bibinfo {year}
  {2024}{\natexlab{b}})},\ \Eprint {http://arxiv.org/abs/2405.09484}
  {arXiv:2405.09484 [hep-lat]} \BibitemShut {NoStop}%
\bibitem [{\citenamefont {Son}\ and\ \citenamefont
  {Yamamoto}(2012)}]{Son:2012wh}%
  \BibitemOpen
  \bibfield  {author} {\bibinfo {author} {\bibfnamefont {D.~T.}\ \bibnamefont
  {Son}}\ and\ \bibinfo {author} {\bibfnamefont {N.}~\bibnamefont {Yamamoto}},\
  }\href {\doibase 10.1103/PhysRevLett.109.181602} {\bibfield  {journal}
  {\bibinfo  {journal} {Phys. Rev. Lett.}\ }\textbf {\bibinfo {volume} {109}},\
  \bibinfo {pages} {181602} (\bibinfo {year} {2012})},\ \Eprint
  {http://arxiv.org/abs/1203.2697} {arXiv:1203.2697 [cond-mat.mes-hall]}
  \BibitemShut {NoStop}%
\bibitem [{\citenamefont {Chen}\ \emph {et~al.}(2013)\citenamefont {Chen},
  \citenamefont {Pu}, \citenamefont {Wang},\ and\ \citenamefont
  {Wang}}]{Chen:2012ca}%
  \BibitemOpen
  \bibfield  {author} {\bibinfo {author} {\bibfnamefont {J.-W.}\ \bibnamefont
  {Chen}}, \bibinfo {author} {\bibfnamefont {S.}~\bibnamefont {Pu}}, \bibinfo
  {author} {\bibfnamefont {Q.}~\bibnamefont {Wang}}, \ and\ \bibinfo {author}
  {\bibfnamefont {X.-N.}\ \bibnamefont {Wang}},\ }\href {\doibase
  10.1103/PhysRevLett.110.262301} {\bibfield  {journal} {\bibinfo  {journal}
  {Phys. Rev. Lett.}\ }\textbf {\bibinfo {volume} {110}},\ \bibinfo {pages}
  {262301} (\bibinfo {year} {2013})},\ \Eprint {http://arxiv.org/abs/1210.8312}
  {arXiv:1210.8312 [hep-th]} \BibitemShut {NoStop}%
\bibitem [{\citenamefont {Chen}\ \emph {et~al.}(2014)\citenamefont {Chen},
  \citenamefont {Son}, \citenamefont {Stephanov}, \citenamefont {Yee},\ and\
  \citenamefont {Yin}}]{Chen:2014cla}%
  \BibitemOpen
  \bibfield  {author} {\bibinfo {author} {\bibfnamefont {J.-Y.}\ \bibnamefont
  {Chen}}, \bibinfo {author} {\bibfnamefont {D.~T.}\ \bibnamefont {Son}},
  \bibinfo {author} {\bibfnamefont {M.~A.}\ \bibnamefont {Stephanov}}, \bibinfo
  {author} {\bibfnamefont {H.-U.}\ \bibnamefont {Yee}}, \ and\ \bibinfo
  {author} {\bibfnamefont {Y.}~\bibnamefont {Yin}},\ }\href {\doibase
  10.1103/PhysRevLett.113.182302} {\bibfield  {journal} {\bibinfo  {journal}
  {Phys. Rev. Lett.}\ }\textbf {\bibinfo {volume} {113}},\ \bibinfo {pages}
  {182302} (\bibinfo {year} {2014})},\ \Eprint {http://arxiv.org/abs/1404.5963}
  {arXiv:1404.5963 [hep-th]} \BibitemShut {NoStop}%
\bibitem [{\citenamefont {Li}\ \emph {et~al.}(2016)\citenamefont {Li},
  \citenamefont {Kharzeev}, \citenamefont {Zhang}, \citenamefont {Huang},
  \citenamefont {Pletikosic}, \citenamefont {Fedorov}, \citenamefont {Zhong},
  \citenamefont {Schneeloch}, \citenamefont {Gu},\ and\ \citenamefont
  {Valla}}]{Li:2014bha}%
  \BibitemOpen
  \bibfield  {author} {\bibinfo {author} {\bibfnamefont {Q.}~\bibnamefont
  {Li}}, \bibinfo {author} {\bibfnamefont {D.~E.}\ \bibnamefont {Kharzeev}},
  \bibinfo {author} {\bibfnamefont {C.}~\bibnamefont {Zhang}}, \bibinfo
  {author} {\bibfnamefont {Y.}~\bibnamefont {Huang}}, \bibinfo {author}
  {\bibfnamefont {I.}~\bibnamefont {Pletikosic}}, \bibinfo {author}
  {\bibfnamefont {A.~V.}\ \bibnamefont {Fedorov}}, \bibinfo {author}
  {\bibfnamefont {R.~D.}\ \bibnamefont {Zhong}}, \bibinfo {author}
  {\bibfnamefont {J.~A.}\ \bibnamefont {Schneeloch}}, \bibinfo {author}
  {\bibfnamefont {G.~D.}\ \bibnamefont {Gu}}, \ and\ \bibinfo {author}
  {\bibfnamefont {T.}~\bibnamefont {Valla}},\ }\href {\doibase
  10.1038/nphys3648} {\bibfield  {journal} {\bibinfo  {journal} {Nature Phys.}\
  }\textbf {\bibinfo {volume} {12}},\ \bibinfo {pages} {550} (\bibinfo {year}
  {2016})},\ \Eprint {http://arxiv.org/abs/1412.6543} {arXiv:1412.6543
  [cond-mat.str-el]} \BibitemShut {NoStop}%
\bibitem [{\citenamefont {Huang}(2016)}]{Huang:2015oca}%
  \BibitemOpen
  \bibfield  {author} {\bibinfo {author} {\bibfnamefont {X.-G.}\ \bibnamefont
  {Huang}},\ }\href {\doibase 10.1088/0034-4885/79/7/076302} {\bibfield
  {journal} {\bibinfo  {journal} {Rept. Prog. Phys.}\ }\textbf {\bibinfo
  {volume} {79}},\ \bibinfo {pages} {076302} (\bibinfo {year} {2016})},\
  \Eprint {http://arxiv.org/abs/1509.04073} {arXiv:1509.04073 [nucl-th]}
  \BibitemShut {NoStop}%
\bibitem [{\citenamefont {Kharzeev}\ and\ \citenamefont
  {Liao}(2021)}]{Kharzeev:2020jxw}%
  \BibitemOpen
  \bibfield  {author} {\bibinfo {author} {\bibfnamefont {D.~E.}\ \bibnamefont
  {Kharzeev}}\ and\ \bibinfo {author} {\bibfnamefont {J.}~\bibnamefont
  {Liao}},\ }\href {\doibase 10.1038/s42254-020-00254-6} {\bibfield  {journal}
  {\bibinfo  {journal} {Nature Rev. Phys.}\ }\textbf {\bibinfo {volume} {3}},\
  \bibinfo {pages} {55} (\bibinfo {year} {2021})},\ \Eprint
  {http://arxiv.org/abs/2102.06623} {arXiv:2102.06623 [hep-ph]} \BibitemShut
  {NoStop}%
\bibitem [{\citenamefont {Adler}(1969)}]{Adler:1969gk}%
  \BibitemOpen
  \bibfield  {author} {\bibinfo {author} {\bibfnamefont {S.~L.}\ \bibnamefont
  {Adler}},\ }\href {\doibase 10.1103/PhysRev.177.2426} {\bibfield  {journal}
  {\bibinfo  {journal} {Phys. Rev.}\ }\textbf {\bibinfo {volume} {177}},\
  \bibinfo {pages} {2426} (\bibinfo {year} {1969})}\BibitemShut {NoStop}%
\bibitem [{\citenamefont {Bell}\ and\ \citenamefont
  {Jackiw}(1969)}]{Bell:1969ts}%
  \BibitemOpen
  \bibfield  {author} {\bibinfo {author} {\bibfnamefont {J.~S.}\ \bibnamefont
  {Bell}}\ and\ \bibinfo {author} {\bibfnamefont {R.}~\bibnamefont {Jackiw}},\
  }\href {\doibase 10.1007/BF02823296} {\bibfield  {journal} {\bibinfo
  {journal} {Nuovo Cim. A}\ }\textbf {\bibinfo {volume} {60}},\ \bibinfo
  {pages} {47} (\bibinfo {year} {1969})}\BibitemShut {NoStop}%
\bibitem [{\citenamefont {Bardeen}(1969)}]{Bardeen:1969md}%
  \BibitemOpen
  \bibfield  {author} {\bibinfo {author} {\bibfnamefont {W.~A.}\ \bibnamefont
  {Bardeen}},\ }\href {\doibase 10.1103/PhysRev.184.1848} {\bibfield  {journal}
  {\bibinfo  {journal} {Phys. Rev.}\ }\textbf {\bibinfo {volume} {184}},\
  \bibinfo {pages} {1848} (\bibinfo {year} {1969})}\BibitemShut {NoStop}%
\bibitem [{\citenamefont {Fujikawa}(1985)}]{Fujikawa:1984pt}%
  \BibitemOpen
  \bibfield  {author} {\bibinfo {author} {\bibfnamefont {K.}~\bibnamefont
  {Fujikawa}},\ }\href {\doibase 10.1103/PhysRevD.31.341} {\bibfield  {journal}
  {\bibinfo  {journal} {Phys. Rev. D}\ }\textbf {\bibinfo {volume} {31}},\
  \bibinfo {pages} {341} (\bibinfo {year} {1985})}\BibitemShut {NoStop}%
\bibitem [{\citenamefont {Kondo}(1997)}]{Kondo:1996xn}%
  \BibitemOpen
  \bibfield  {author} {\bibinfo {author} {\bibfnamefont {K.-I.}\ \bibnamefont
  {Kondo}},\ }\href {\doibase 10.1142/S0217751X97002978} {\bibfield  {journal}
  {\bibinfo  {journal} {Int. J. Mod. Phys. A}\ }\textbf {\bibinfo {volume}
  {12}},\ \bibinfo {pages} {5651} (\bibinfo {year} {1997})},\ \Eprint
  {http://arxiv.org/abs/hep-th/9608100} {arXiv:hep-th/9608100} \BibitemShut
  {NoStop}%
\bibitem [{\citenamefont {He}\ \emph {et~al.}(2000)\citenamefont {He},
  \citenamefont {Khanna},\ and\ \citenamefont {Takahashi}}]{He:2000we}%
  \BibitemOpen
  \bibfield  {author} {\bibinfo {author} {\bibfnamefont {H.-X.}\ \bibnamefont
  {He}}, \bibinfo {author} {\bibfnamefont {F.~C.}\ \bibnamefont {Khanna}}, \
  and\ \bibinfo {author} {\bibfnamefont {Y.}~\bibnamefont {Takahashi}},\ }\href
  {\doibase 10.1016/S0370-2693(00)00353-1} {\bibfield  {journal} {\bibinfo
  {journal} {Phys. Lett. B}\ }\textbf {\bibinfo {volume} {480}},\ \bibinfo
  {pages} {222} (\bibinfo {year} {2000})}\BibitemShut {NoStop}%
\bibitem [{\citenamefont {Qin}\ \emph {et~al.}(2013)\citenamefont {Qin},
  \citenamefont {Chang}, \citenamefont {Liu}, \citenamefont {Roberts},\ and\
  \citenamefont {Schmidt}}]{Qin:2013mta}%
  \BibitemOpen
  \bibfield  {author} {\bibinfo {author} {\bibfnamefont {S.-X.}\ \bibnamefont
  {Qin}}, \bibinfo {author} {\bibfnamefont {L.}~\bibnamefont {Chang}}, \bibinfo
  {author} {\bibfnamefont {Y.-X.}\ \bibnamefont {Liu}}, \bibinfo {author}
  {\bibfnamefont {C.~D.}\ \bibnamefont {Roberts}}, \ and\ \bibinfo {author}
  {\bibfnamefont {S.~M.}\ \bibnamefont {Schmidt}},\ }\href {\doibase
  10.1016/j.physletb.2013.04.034} {\bibfield  {journal} {\bibinfo  {journal}
  {Phys. Lett. B}\ }\textbf {\bibinfo {volume} {722}},\ \bibinfo {pages} {384}
  (\bibinfo {year} {2013})},\ \Eprint {http://arxiv.org/abs/1302.3276}
  {arXiv:1302.3276 [nucl-th]} \BibitemShut {NoStop}%
\bibitem [{\citenamefont {Gynther}\ \emph {et~al.}(2011)\citenamefont
  {Gynther}, \citenamefont {Landsteiner}, \citenamefont {Pena-Benitez},\ and\
  \citenamefont {Rebhan}}]{Gynther:2010ed}%
  \BibitemOpen
  \bibfield  {author} {\bibinfo {author} {\bibfnamefont {A.}~\bibnamefont
  {Gynther}}, \bibinfo {author} {\bibfnamefont {K.}~\bibnamefont
  {Landsteiner}}, \bibinfo {author} {\bibfnamefont {F.}~\bibnamefont
  {Pena-Benitez}}, \ and\ \bibinfo {author} {\bibfnamefont {A.}~\bibnamefont
  {Rebhan}},\ }\href {\doibase 10.1007/JHEP02(2011)110} {\bibfield  {journal}
  {\bibinfo  {journal} {JHEP}\ }\textbf {\bibinfo {volume} {02}},\ \bibinfo
  {pages} {110} (\bibinfo {year} {2011})},\ \Eprint
  {http://arxiv.org/abs/1005.2587} {arXiv:1005.2587 [hep-th]} \BibitemShut
  {NoStop}%
\bibitem [{\citenamefont {Hou}\ \emph {et~al.}(2011)\citenamefont {Hou},
  \citenamefont {Liu},\ and\ \citenamefont {Ren}}]{Hou:2011ze}%
  \BibitemOpen
  \bibfield  {author} {\bibinfo {author} {\bibfnamefont {D.}~\bibnamefont
  {Hou}}, \bibinfo {author} {\bibfnamefont {H.}~\bibnamefont {Liu}}, \ and\
  \bibinfo {author} {\bibfnamefont {H.-c.}\ \bibnamefont {Ren}},\ }\href
  {\doibase 10.1007/JHEP05(2011)046} {\bibfield  {journal} {\bibinfo  {journal}
  {JHEP}\ }\textbf {\bibinfo {volume} {05}},\ \bibinfo {pages} {046} (\bibinfo
  {year} {2011})},\ \Eprint {http://arxiv.org/abs/1103.2035} {arXiv:1103.2035
  [hep-ph]} \BibitemShut {NoStop}%
\bibitem [{\citenamefont {Gao}\ \emph {et~al.}(2012)\citenamefont {Gao},
  \citenamefont {Liang}, \citenamefont {Pu}, \citenamefont {Wang},\ and\
  \citenamefont {Wang}}]{Gao:2012ix}%
  \BibitemOpen
  \bibfield  {author} {\bibinfo {author} {\bibfnamefont {J.-H.}\ \bibnamefont
  {Gao}}, \bibinfo {author} {\bibfnamefont {Z.-T.}\ \bibnamefont {Liang}},
  \bibinfo {author} {\bibfnamefont {S.}~\bibnamefont {Pu}}, \bibinfo {author}
  {\bibfnamefont {Q.}~\bibnamefont {Wang}}, \ and\ \bibinfo {author}
  {\bibfnamefont {X.-N.}\ \bibnamefont {Wang}},\ }\href {\doibase
  10.1103/PhysRevLett.109.232301} {\bibfield  {journal} {\bibinfo  {journal}
  {Phys. Rev. Lett.}\ }\textbf {\bibinfo {volume} {109}},\ \bibinfo {pages}
  {232301} (\bibinfo {year} {2012})},\ \Eprint {http://arxiv.org/abs/1203.0725}
  {arXiv:1203.0725 [hep-ph]} \BibitemShut {NoStop}%
\bibitem [{\citenamefont {Wu}\ \emph {et~al.}(2017)\citenamefont {Wu},
  \citenamefont {Hou},\ and\ \citenamefont {Ren}}]{Wu:2016dam}%
  \BibitemOpen
  \bibfield  {author} {\bibinfo {author} {\bibfnamefont {Y.}~\bibnamefont
  {Wu}}, \bibinfo {author} {\bibfnamefont {D.}~\bibnamefont {Hou}}, \ and\
  \bibinfo {author} {\bibfnamefont {H.-c.}\ \bibnamefont {Ren}},\ }\href
  {\doibase 10.1103/PhysRevD.96.096015} {\bibfield  {journal} {\bibinfo
  {journal} {Phys. Rev. D}\ }\textbf {\bibinfo {volume} {96}},\ \bibinfo
  {pages} {096015} (\bibinfo {year} {2017})},\ \Eprint
  {http://arxiv.org/abs/1601.06520} {arXiv:1601.06520 [hep-ph]} \BibitemShut
  {NoStop}%
\bibitem [{\citenamefont {Hattori}\ and\ \citenamefont
  {Huang}(2017)}]{Hattori:2016emy}%
  \BibitemOpen
  \bibfield  {author} {\bibinfo {author} {\bibfnamefont {K.}~\bibnamefont
  {Hattori}}\ and\ \bibinfo {author} {\bibfnamefont {X.-G.}\ \bibnamefont
  {Huang}},\ }\href {\doibase 10.1007/s41365-016-0178-3} {\bibfield  {journal}
  {\bibinfo  {journal} {Nucl. Sci. Tech.}\ }\textbf {\bibinfo {volume} {28}},\
  \bibinfo {pages} {26} (\bibinfo {year} {2017})},\ \Eprint
  {http://arxiv.org/abs/1609.00747} {arXiv:1609.00747 [nucl-th]} \BibitemShut
  {NoStop}%
\bibitem [{\citenamefont {Deng}\ \emph {et~al.}(2016)\citenamefont {Deng},
  \citenamefont {Huang}, \citenamefont {Ma},\ and\ \citenamefont
  {Wang}}]{Deng:2016knn}%
  \BibitemOpen
  \bibfield  {author} {\bibinfo {author} {\bibfnamefont {W.-T.}\ \bibnamefont
  {Deng}}, \bibinfo {author} {\bibfnamefont {X.-G.}\ \bibnamefont {Huang}},
  \bibinfo {author} {\bibfnamefont {G.-L.}\ \bibnamefont {Ma}}, \ and\ \bibinfo
  {author} {\bibfnamefont {G.}~\bibnamefont {Wang}},\ }\href {\doibase
  10.1103/PhysRevC.94.041901} {\bibfield  {journal} {\bibinfo  {journal} {Phys.
  Rev. C}\ }\textbf {\bibinfo {volume} {94}},\ \bibinfo {pages} {041901}
  (\bibinfo {year} {2016})},\ \Eprint {http://arxiv.org/abs/1607.04697}
  {arXiv:1607.04697 [nucl-th]} \BibitemShut {NoStop}%
\bibitem [{\citenamefont {Deng}\ \emph {et~al.}(2018)\citenamefont {Deng},
  \citenamefont {Huang}, \citenamefont {Ma},\ and\ \citenamefont
  {Wang}}]{Deng:2018dut}%
  \BibitemOpen
  \bibfield  {author} {\bibinfo {author} {\bibfnamefont {W.-T.}\ \bibnamefont
  {Deng}}, \bibinfo {author} {\bibfnamefont {X.-G.}\ \bibnamefont {Huang}},
  \bibinfo {author} {\bibfnamefont {G.-L.}\ \bibnamefont {Ma}}, \ and\ \bibinfo
  {author} {\bibfnamefont {G.}~\bibnamefont {Wang}},\ }\href {\doibase
  10.1103/PhysRevC.97.044901} {\bibfield  {journal} {\bibinfo  {journal} {Phys.
  Rev. C}\ }\textbf {\bibinfo {volume} {97}},\ \bibinfo {pages} {044901}
  (\bibinfo {year} {2018})},\ \Eprint {http://arxiv.org/abs/1802.02292}
  {arXiv:1802.02292 [nucl-th]} \BibitemShut {NoStop}%
\bibitem [{\citenamefont {Zhao}\ and\ \citenamefont {Ma}(2022)}]{Zhao:2022grq}%
  \BibitemOpen
  \bibfield  {author} {\bibinfo {author} {\bibfnamefont {X.-L.}\ \bibnamefont
  {Zhao}}\ and\ \bibinfo {author} {\bibfnamefont {G.-L.}\ \bibnamefont {Ma}},\
  }\href {\doibase 10.1103/PhysRevC.106.034909} {\bibfield  {journal} {\bibinfo
   {journal} {Phys. Rev. C}\ }\textbf {\bibinfo {volume} {106}},\ \bibinfo
  {pages} {034909} (\bibinfo {year} {2022})},\ \Eprint
  {http://arxiv.org/abs/2203.15214} {arXiv:2203.15214 [nucl-th]} \BibitemShut
  {NoStop}%
\bibitem [{\citenamefont {Huang}\ \emph {et~al.}(2018)\citenamefont {Huang},
  \citenamefont {Shi}, \citenamefont {Jiang}, \citenamefont {Liao},\ and\
  \citenamefont {Zhuang}}]{Huang:2018wdl}%
  \BibitemOpen
  \bibfield  {author} {\bibinfo {author} {\bibfnamefont {A.}~\bibnamefont
  {Huang}}, \bibinfo {author} {\bibfnamefont {S.}~\bibnamefont {Shi}}, \bibinfo
  {author} {\bibfnamefont {Y.}~\bibnamefont {Jiang}}, \bibinfo {author}
  {\bibfnamefont {J.}~\bibnamefont {Liao}}, \ and\ \bibinfo {author}
  {\bibfnamefont {P.}~\bibnamefont {Zhuang}},\ }\href {\doibase
  10.1103/PhysRevD.98.036010} {\bibfield  {journal} {\bibinfo  {journal} {Phys.
  Rev. D}\ }\textbf {\bibinfo {volume} {98}},\ \bibinfo {pages} {036010}
  (\bibinfo {year} {2018})},\ \Eprint {http://arxiv.org/abs/1801.03640}
  {arXiv:1801.03640 [hep-th]} \BibitemShut {NoStop}%
\bibitem [{\citenamefont {Shi}\ \emph {et~al.}(2020)\citenamefont {Shi},
  \citenamefont {Zhang}, \citenamefont {Hou},\ and\ \citenamefont
  {Liao}}]{Shi:2019wzi}%
  \BibitemOpen
  \bibfield  {author} {\bibinfo {author} {\bibfnamefont {S.}~\bibnamefont
  {Shi}}, \bibinfo {author} {\bibfnamefont {H.}~\bibnamefont {Zhang}}, \bibinfo
  {author} {\bibfnamefont {D.}~\bibnamefont {Hou}}, \ and\ \bibinfo {author}
  {\bibfnamefont {J.}~\bibnamefont {Liao}},\ }\href {\doibase
  10.1103/PhysRevLett.125.242301} {\bibfield  {journal} {\bibinfo  {journal}
  {Phys. Rev. Lett.}\ }\textbf {\bibinfo {volume} {125}},\ \bibinfo {pages}
  {242301} (\bibinfo {year} {2020})},\ \Eprint
  {http://arxiv.org/abs/1910.14010} {arXiv:1910.14010 [nucl-th]} \BibitemShut
  {NoStop}%
\bibitem [{\citenamefont {Kharzeev}\ \emph {et~al.}(2022)\citenamefont
  {Kharzeev}, \citenamefont {Liao},\ and\ \citenamefont
  {Shi}}]{Kharzeev:2022hqz}%
  \BibitemOpen
  \bibfield  {author} {\bibinfo {author} {\bibfnamefont {D.~E.}\ \bibnamefont
  {Kharzeev}}, \bibinfo {author} {\bibfnamefont {J.}~\bibnamefont {Liao}}, \
  and\ \bibinfo {author} {\bibfnamefont {S.}~\bibnamefont {Shi}},\ }\href
  {\doibase 10.1103/PhysRevC.106.L051903} {\bibfield  {journal} {\bibinfo
  {journal} {Phys. Rev. C}\ }\textbf {\bibinfo {volume} {106}},\ \bibinfo
  {pages} {L051903} (\bibinfo {year} {2022})},\ \Eprint
  {http://arxiv.org/abs/2205.00120} {arXiv:2205.00120 [nucl-th]} \BibitemShut
  {NoStop}%
\bibitem [{\citenamefont {Ferrer}\ and\ \citenamefont {de~la
  Incera}(2009)}]{Ferrer:2008dy}%
  \BibitemOpen
  \bibfield  {author} {\bibinfo {author} {\bibfnamefont {E.~J.}\ \bibnamefont
  {Ferrer}}\ and\ \bibinfo {author} {\bibfnamefont {V.}~\bibnamefont {de~la
  Incera}},\ }\href {\doibase 10.1103/PhysRevLett.102.050402} {\bibfield
  {journal} {\bibinfo  {journal} {Phys. Rev. Lett.}\ }\textbf {\bibinfo
  {volume} {102}},\ \bibinfo {pages} {050402} (\bibinfo {year} {2009})},\
  \Eprint {http://arxiv.org/abs/0807.4744} {arXiv:0807.4744 [hep-ph]}
  \BibitemShut {NoStop}%
\bibitem [{\citenamefont {Watson}\ and\ \citenamefont
  {Reinhardt}(2014)}]{Watson:2013ghq}%
  \BibitemOpen
  \bibfield  {author} {\bibinfo {author} {\bibfnamefont {P.}~\bibnamefont
  {Watson}}\ and\ \bibinfo {author} {\bibfnamefont {H.}~\bibnamefont
  {Reinhardt}},\ }\href {\doibase 10.1103/PhysRevD.89.045008} {\bibfield
  {journal} {\bibinfo  {journal} {Phys. Rev. D}\ }\textbf {\bibinfo {volume}
  {89}},\ \bibinfo {pages} {045008} (\bibinfo {year} {2014})},\ \Eprint
  {http://arxiv.org/abs/1310.6050} {arXiv:1310.6050 [hep-ph]} \BibitemShut
  {NoStop}%
\bibitem [{\citenamefont {Ding}\ \emph {et~al.}(2025)\citenamefont {Ding},
  \citenamefont {Gao},\ and\ \citenamefont {Schmidt}}]{Ding:2025zqu}%
  \BibitemOpen
  \bibfield  {author} {\bibinfo {author} {\bibfnamefont {M.}~\bibnamefont
  {Ding}}, \bibinfo {author} {\bibfnamefont {F.}~\bibnamefont {Gao}}, \ and\
  \bibinfo {author} {\bibfnamefont {S.~M.}\ \bibnamefont {Schmidt}},\
  }\href@noop {} {\  (\bibinfo {year} {2025})},\ \Eprint
  {http://arxiv.org/abs/2504.14504} {arXiv:2504.14504 [hep-ph]} \BibitemShut
  {NoStop}%
\bibitem [{\citenamefont {Ferrer}\ and\ \citenamefont {de~la
  Incera}(2010)}]{Ferrer:2009nq}%
  \BibitemOpen
  \bibfield  {author} {\bibinfo {author} {\bibfnamefont {E.~J.}\ \bibnamefont
  {Ferrer}}\ and\ \bibinfo {author} {\bibfnamefont {V.}~\bibnamefont {de~la
  Incera}},\ }\href {\doibase 10.1016/j.nuclphysb.2009.08.024} {\bibfield
  {journal} {\bibinfo  {journal} {Nucl. Phys. B}\ }\textbf {\bibinfo {volume}
  {824}},\ \bibinfo {pages} {217} (\bibinfo {year} {2010})},\ \Eprint
  {http://arxiv.org/abs/0905.1733} {arXiv:0905.1733 [hep-ph]} \BibitemShut
  {NoStop}%
\bibitem [{\citenamefont {Goldstone}\ and\ \citenamefont
  {Wilczek}(1981)}]{Goldstone:1981kk}%
  \BibitemOpen
  \bibfield  {author} {\bibinfo {author} {\bibfnamefont {J.}~\bibnamefont
  {Goldstone}}\ and\ \bibinfo {author} {\bibfnamefont {F.}~\bibnamefont
  {Wilczek}},\ }\href {\doibase 10.1103/PhysRevLett.47.986} {\bibfield
  {journal} {\bibinfo  {journal} {Phys. Rev. Lett.}\ }\textbf {\bibinfo
  {volume} {47}},\ \bibinfo {pages} {986} (\bibinfo {year} {1981})}\BibitemShut
  {NoStop}%
\bibitem [{\citenamefont {Son}\ and\ \citenamefont
  {Stephanov}(2008)}]{Son:2007ny}%
  \BibitemOpen
  \bibfield  {author} {\bibinfo {author} {\bibfnamefont {D.~T.}\ \bibnamefont
  {Son}}\ and\ \bibinfo {author} {\bibfnamefont {M.~A.}\ \bibnamefont
  {Stephanov}},\ }\href {\doibase 10.1103/PhysRevD.77.014021} {\bibfield
  {journal} {\bibinfo  {journal} {Phys. Rev. D}\ }\textbf {\bibinfo {volume}
  {77}},\ \bibinfo {pages} {014021} (\bibinfo {year} {2008})},\ \Eprint
  {http://arxiv.org/abs/0710.1084} {arXiv:0710.1084 [hep-ph]} \BibitemShut
  {NoStop}%
\end{thebibliography}%

\appendix

\section{Dressed  quark  propagator in the Ritus basis and the quark gap equation in the presence of magnetic field}\label{App:ritus}
 The coefficients of the inverse quark propagator  and   quark propagator  in the presence of magnetic field are  no longer following the conventional inverse relation. In the Ritus basis, their relation can be expressed as 	
\begin{align*}
	W_{Z^+_\parallel}=&\frac{\Delta_+Z^-_\parallel-\Delta_-M^-}{\Delta_{+}^2+p^2_\parallel\Delta_{-}^2}\,,&W_{M^+}=&\frac{\Delta_+M^- -p_\parallel^2Z^-_\parallel\Delta_-}{\Delta_{+}^2+p^2_\parallel\Delta_{-}^2}\,,\\	W_{Z^-_\parallel}=&\frac{\Delta_+Z^+_\parallel+\Delta_-M^-}{\Delta_{+}^2+p^2_\parallel\Delta_{-}^2}\,,&W_{M^-}=&\frac{\Delta_+M^++p_\parallel^2Z^+_\parallel\Delta_-}{\Delta_{+}^2+p^2_\parallel\Delta_{-}^2}\,,\\
	W_{Z_\perp}=&\frac{\Delta_+Z_\perp}{\Delta_{+}^2+p^2_\parallel\Delta_{-}^2}\,.{\addtocounter{equation}{1}\tag{\theequation}}
\end{align*}
	with $$\Delta_+(p_\parallel,n)=p_\parallel^2 Z^+_\parallel Z^-_\parallel+M^+ M^- +2 n |q_f B| Z^2_\perp,$$ and $$\Delta_-(p_\parallel,n)= Z^+_\parallel  M^-  - M^+Z^-_\parallel. $$
For the tree level propagator, one has:
 \begin{equation}
Z^\pm_\parallel=Z_\perp=1,\quad M^{\pm}=m_f, 
\end{equation}
and consequently, the coefficients of the quark propagator become:

\begin{align*}
	W_{Z^\pm_\parallel}(p_\parallel,n)=&W_{Z_\perp}(p_\parallel,n)=\frac{1}{ p^2_\parallel+m_f^2+2 n|q_f B|}\,,\\
	W_{M^\pm}(p_\parallel,n)=&\frac{m_f}{p^2_\parallel+m_f^2+2 n|q_f B|}\,.{\addtocounter{equation}{1}\tag{\theequation}}
\end{align*}

One can  sum up all the orders of  the Hermite function  in  coefficients $W_X$ analytically. To achieve this, the dependence of the Hermite function order in the coefficients $Z^\pm_{\parallel,\perp}$ and $M^\pm$ are neglected, and the coefficients are expressed as:

\begin{eqnarray}\label{eq:propa}
&&W_{ Z^+_\parallel}(p) =\int ds e^{-s\frac{ \Delta^+(p)}{ Z^2_\perp(p)  }-\frac{p_\perp^2}{q_f B}{\rm tanh}[q_f B s ]}\notag\\
&&\times \left\{(1-{\rm tanh}[q_f B s])\frac{Z^-_\parallel(p) }{Z^2_\perp(p)}+s\frac{\Delta^-}{Z^4_\perp(p)} M^-(p)\right\}\,,\notag\\
&&W_{ Z^-_\parallel}(p) =\int ds e^{-s\frac{ \Delta^+(p)}{ Z^2_\perp(p)  }-\frac{p_\perp^2}{q_f B}{\rm tanh}[q_f B s ]}\notag\\
&&\times \left\{(1+{\rm tanh}[q_f B s])\frac{Z^+_\parallel(p) }{Z^2_\perp(p)}-s\frac{\Delta^-}{Z^4_\perp(p)} M^+(p)\right\}\,,\notag\\
&&W_{ Z_\perp}(p) =\int ds e^{-s\frac{ \Delta^+(p)}{ Z^2_\perp(p)  }-\frac{p_\perp^2}{q_f B}{\rm tanh}[q_f B s ]}\notag\\
&&\times \left\{(1-{\rm tanh}^2[q_f B s])\frac{1 }{Z_\perp(p)}\right\}\,,\notag
\end{eqnarray}

\begin{eqnarray}
&&W_{ M^+}(p) =\int ds e^{-s\frac{ \Delta^+(p)}{ Z^2_\perp(p)  }-\frac{p_\perp^2}{q_f B}{\rm tanh}[q_f B s ]}\notag\\
&&\times \left\{(1-{\rm tanh}[q_f B s])\frac{M^+(p) }{Z^2_\perp(p)}-s\frac{\Delta^-}{Z^4_\perp(p)} Z^-_\parallel(p)p^2_\parallel \right\}\,,\notag\\
&&W_{ M^-}(p) =\int ds e^{-s\frac{ \Delta^+(p)}{ Z^2_\perp(p)  }-\frac{p_\perp^2}{q_f B}{\rm tanh}[q_f B s ]}\notag\\
&&\times \left\{(1+{\rm tanh}[q_f B s])\frac{M^+(p) }{Z^2_\perp(p)}+s\frac{\Delta^-}{Z^4_\perp(p)} Z^+_\parallel(p)p^2_\parallel \right\}\,,\notag
	\end{eqnarray}
	with $$\Delta^+(p)=p_\parallel^2 Z^+_\parallel(p)Z^-_\parallel(p)  +M^+(p)M^-(p),$$ and $$\Delta^-(p)= Z^+_\parallel(p)  M^-(p) - M^+(p)  Z^-_\parallel(p). $$

The Schwinger phase  can be completely factored  out  from the gap equation if the interaction kernel contains only the transferred momentum. The factoring out of the Schwinger phase leads to a sophisticated result that  the gap equation  in constant magnetic field  is precisely the same as the conventional gap equation without magnetic field, which reads as:

\begin{align}
{S}^{-1} (p) =S_0(p)+T\sum_{n\in \mathbbm{Z}} \int \frac{d^3 q}{(2 \pi)^3}\frac{4}{3} g_s \gamma_\mu S(q)\Gamma_{\nu}(k) D_{\mu\nu}(k)\,,
		\label{eq:QuarkDSE2}
	\end{align}
with now
 \begin{eqnarray}
{S}^{-1} (p)&&=\Sigma^+(M^+(p)  + Z^+_\parallel(p)  \slashed{p}_\parallel) + Z_\perp(p)  \slashed{p}_\perp \notag\\
&&+\Sigma^-(M^-(p)  + Z^-_\parallel(p)  \slashed{p}_\parallel) ,\\
 S (p)&&=\Sigma^+(W_{M^+}(p)  +W_{ Z^+_\parallel}(p)  \slashed{p}_\parallel) + W_{Z_\perp}(p)  \slashed{p}_\perp \notag\\
&&+\Sigma^-(W_{M^-}(p)  + W_{Z^-_\parallel}(p)  \slashed{p}_\parallel).
		\label{eq:propa}
	\end{eqnarray}
The gap equation is then solved in the conventional minimal QCD scheme and then with the numerical solutions of quark propagator, the CME conductivity can be obtained.
	\end{document}